# Star Classification: A Deep Learning Approach for Identifying Binary and Exoplanet Stars

Aman Kumar,[1] and Sarvesh Gharat [2] ⋆

[1]*Tezpur University, Tezpur, Assam 784028, India.*
[2]*Centre for Machine Intelligence and Data Science, Indian Institute of Technology Bombay, 400076, Mumbai, India*

**ABSTRACT**
We present a novel approach for classifying stars as binary or exoplanet using deep learning techniques. Our method utilizes feature extraction, wavelet transformation, and a neural network on the light curves of stars to achieve high-accuracy results. We have also compiled a dataset of binary and exoplanet stars for training and validation by cross-matching observations from multiple space-based telescopes with catalogs of known binary and exoplanet stars. The application of wavelet transformation on the light curves has reduced the number of data points and improved the training time. Our algorithm has shown exceptional performance, with a test accuracy of 81.17%. This method can be applied to large datasets from current and future space-based telescopes, providing an efficient and accurate way of classifying stars.

## 1 INTRODUCTION

The astronomical data is exploding as time passes and through the various instruments deployed by us, we are generating an enormous amount of data. One such instrument is *TESS*. The Transiting Exoplanet Survey Satellite (TESS) is a mission launched on April 18, 2018, whose goal is to discover earth-sized exoplanets orbiting the brightest dwarf stars Ricker et al. (2015); Barclay et al. (2018). TESS is equipped with four cameras, known as charge-coupled devices (CCDs), that can capture images of a region of the sky measuring 24 degrees by 96 degrees. The cameras will observe the brightness of 15,000 to 20,000 stars every 2 minute for a period of 27 to 356 days Chrisp et al. (2015); Fausnaugh et al. (2021). The data collected by TESS is processed at the Science Processing Operations Center (SPOC) and made available to the public in the form of Target Pixel Files (TPFs) and calibrated light curves Jenkins et al. (2016), with a nominal cadence of 2 minutes for science data and 30 minutes for Full Frame Images (FFIs). TESS also maintains its own catalog of data, mostly observations of luminous objects taken by the satellite.

The Transiting Exoplanet Survey Satellite observes a variety of objects in the sky, ranging from stars to galaxies Muirhead et al. (2018); Guerrero et al. (2021). However, there is a need for rapid identification Yu et al. (2019b) of these objects in the data as it is being generated. The primary target of observation for the TESS mission is stars, particularly host stars. The goal is to detect transits, which are the passing of planets between their host stars and us. These transits block a small amount of the star's light as they pass by, thereby helping us to detect exoplanets. However, similar transits are also observed in the case of eclipsing binaries. As discussed in Ricker et al. (2015); Sullivan et al. (2017), this can result in a large number of false positives, the majority of which are eclipsing binaries, which is a problem that needs to be addressed. Hence, classifying light curves that show transits into host stars and eclipsing binaries is a crucial task in astronomical studies, as both eclipsing binary systems and host stars are important objects Jones et al. (2020); Prša et al. (2022b); Čokina et al. (2021); Howell et al. (2021); Campante et al. (2016); Kane et al. (2020); Campante et al. (2016) of study. The problem of classifying these time series data into binary categories is known as binary time series classification. Rapid identification of these objects is crucial in order to effectively analyze and interpret the data collected by TESS and other similar missions.

Given the recent trend of using machine learning to streamline the process of classification, we decide to employ a deep-learning model to classify the light curves in question. There have been several works in various fields that have used various deep learning models to classify time series data. For example, Xu et al. (2019) uses a Long Short-term Memory (LSTM) based deep learning model to classify solar radio spectrum, and Jamal & Bloom (2020) uses deep learning to classify variable stars. These studies demonstrate the potential of using deep learning models for time series classification and inspired us to investigate the use of deep learning for classifying TESS light curves. In addition to deep learning-based approaches, some works like Dubath et al. (2011) have explored the use of traditional machine learning-based classification methods for similar tasks. These studies use various parameters, such as the period, amplitude, V-I color index, and absolute magnitude, to classify periodic variable stars using techniques like random forests. In summary, we present a method for classifying host stars and eclipsing binaries observed by TESS with high accuracy. This method utilizes both statistical parameters of the light curves and wavelet-transformed light curves as inputs to a deep learning model.

⋆ sarveshgharat19@gmail.com

A wavelet transform, as described in literature such as Chun-Lin (2010) and Zhang & Zhang (2019), is a powerful mathematical tool utilized to analyze signals, such as time series data, in both the time and frequency domains Rhif et al. (2019). The technique is based on the wavelet, which is a small, wave-like function that can be scaled and translated. The wavelet transform decomposes a signal into a series of wavelets, each of which represents a different frequency or scale. The resulting coefficients can be used to represent the signal in a more compact and efficient manner, and can also reveal features of the signal that may be difficult to detect in the original time series data. The use of wavelet transform in dimensionality reduction has been widely studied in the literature, as demonstrated in works such as Keogh et al. (2001) and Ukwatta & Wozniak (2015). The process involves using a technique such as Discrete Wavelet Transform (DWT) Osadchiy et al. (2021); Freire et al. (2019); Al-Qerem et al. (2020) to decompose a signal into a series of detailed and approximation coefficients. The approximation coefficients are used to describe the overall trend of the signal, while the detailed coefficients provide information on the fine details of the signal. By keeping only the approximation coefficients and discarding the detailed coefficients, dimensionality reduction can be achieved while preserving the important features and variability of the original data.

Additionally, several works such as Keogh et al. (2001); Chakrabarti et al. (2002) and Ukwatta & Wozniak (2015) have utilized the Discrete Wavelet Transform (DWT) for dimensionality reduction. This approach is effective in preserving the important features and inherent variability present in the signal while reducing the number of dimensions in the dataset. Therefore, in this context, DWT can be used to effectively reduce the dimensionality of light curves.

## 2 METHODOLOGY

### 2.1 Data Collection and Pre-Processing

The method being used for data processing and cleaning is similar to the one described in the paper "Exploring the TESS Eclipsing Binary Stars" by Pininti et al. (2023). The data sources being used are the VizieR Online Data Catalog: TESS Eclipsing Binary Stars Prsa et al. (2022a) and Exoplanet host stars Su et al. (2022). The data is being cross-checked for observation through TESS and the associated FITS photo is being matched with the GAIA Smart et al. (2021); Prusti et al. (2016); Brown et al. (2018) for validation of objects. This study uses 2-minute cadence SAP fluxes without any preprocessing. This approach is intended to make the study reliable for fast and rapid classification of objects without the need for extensive preprocessing and also to preserve the inherent features of the light curves. The data set includes 804 Host Star light curves and 787 Eclipsing Binary light curves, which have been cleaned for flagged data points and outliers beyond the 3 sigma range. It is important to note that the light curves have not been binned or folded, they were prepared in their original form for training the model. **Furthermore, the durations of the Host Stars and Eclipsing Binaries' periods exhibit a consistent similarity, with the majority of samples having periods of less than 100 days,** **although some exceed this threshold. The distribution of periods for both categories is depicted in Figure 1**

Gaps in the TESS light curves are present due to various reasons, including spacecraft movement and telescope realignment. These gaps, which are random in nature and appear in different positions on various light curves, can be classified as systematic errors in the data Pininti et al. (2023). To address this issue, we employed a method of stitching the light curves on both sides of the gap, effectively filling in the missing data and making the light curve more continuous Hattori et al. (2022), thus improving the performance of our model. **The approach utilized for stitching the light curves in TESS data, which contain gaps, does not involve a complex algorithm. Rather, it involves the removal of the gaps from the light curves by eliminating data points with zero or null values. This process results in the creation of continuous or "stitched" light curves. Two alternative methods were considered to handle this issue. The first approach, implemented in this study, focused on removing the data points associated with unknown flux values, thereby eliminating the gaps. The second approach involved filling these data points with interpolated values. However, the latter method introduces biases based on initial assumptions associated with each interpolation technique. Consequently, we opted to directly remove the gaps instead of filling them with interpolated values, in order to mitigate these biases.**
**To ensure dataset consistency, the light curves were standardized to a length of $12,901$ data points, which corresponded to the shortest light curve within the dataset. This standardization process involved removing data points from the end of longer light curves. Although this resulted in a minor data loss, it was deemed acceptable given that the longest light curve in the dataset contained only 16,007 data points. Standardizing the length of the light curves is crucial for LSTM networks, as they require fixed input lengths. Thus, all inputted light curves must be of the same length.**
**Two approaches were considered to address this issue. The first approach involved trimming the longer light curves to match the length of the shortest light curve. This method incurred a small loss of information, as previously discussed, but the trimmed sections were relatively small. The second approach entailed padding the shorter light curves with a constant value to match the length of the longest light curves. However, this method introduced a bias when selecting the fixed value for padding. After careful consideration of both approaches, we decided to avoid introducing any biases into the data and opted for the light curve standardization through trimming to a fixed length.**
**Subsequent to the previous procedure, the obtained light curves underwent a Discrete Wavelet Transform (DWT) to reduce their dimensions. In this transformation, we specifically focused on the approximation coefficients, which represent the low-pass filtered component of the original light curve. These coefficients offer a smoothed or less detailed version of the initial signal, as they remove high-frequency components through the application of a low-pass filter. Consequently, the resulting signal is smoother and primarily contains low-frequency information, while the high-frequency details are eliminated Pathak (2009) Sifuzzaman et al. (2009). The transformed light curves, which comprised the approximation coefficients, consisted of 6,455 data points. To perform the DWT, we utilized the PyWavelet Python package Lee et al. (2019), employing the "sym5" wavelet. In Figure 3 and Figure 4, we present sample light curves from both**

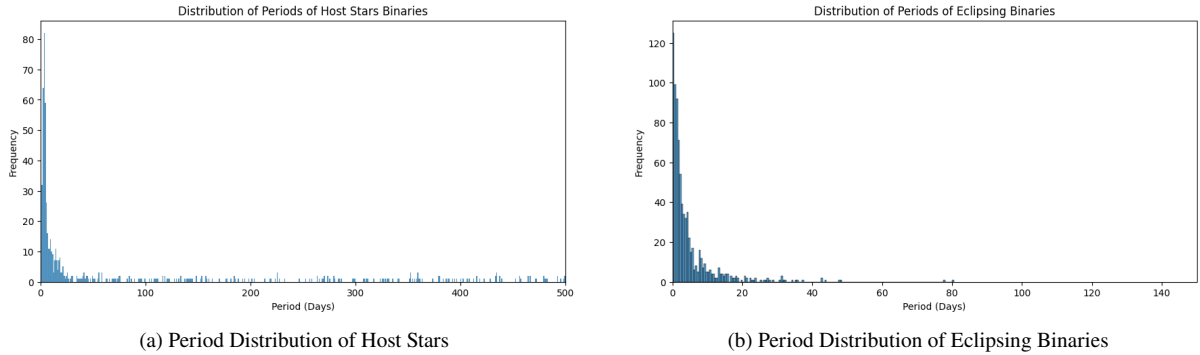

**Figure 1.** Period Distribution of the Host Stars and Eclipsing Binaries present

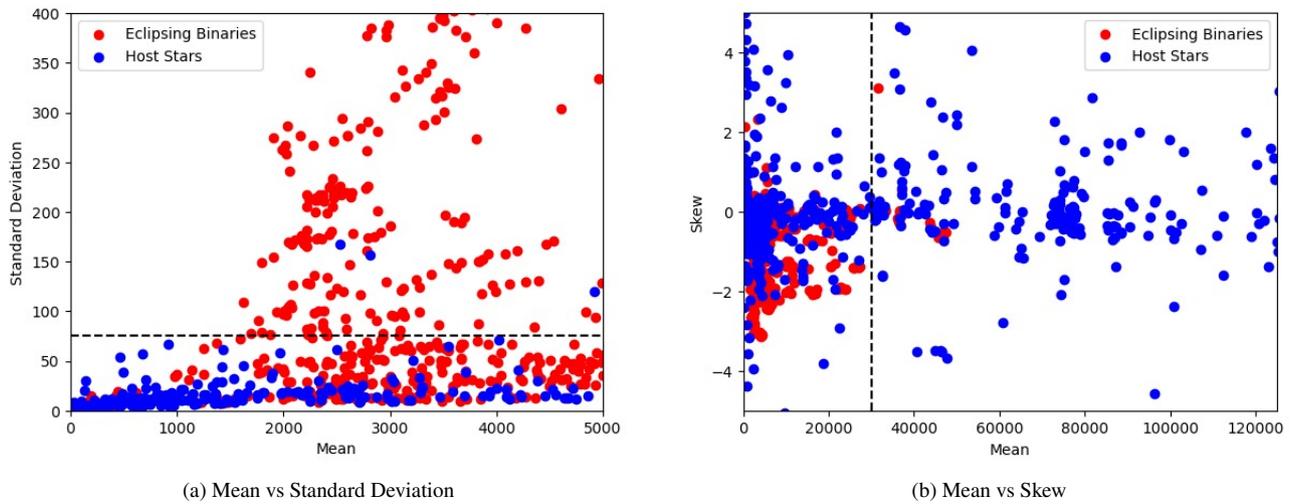

**Figure 2.** Parameter Space Plots

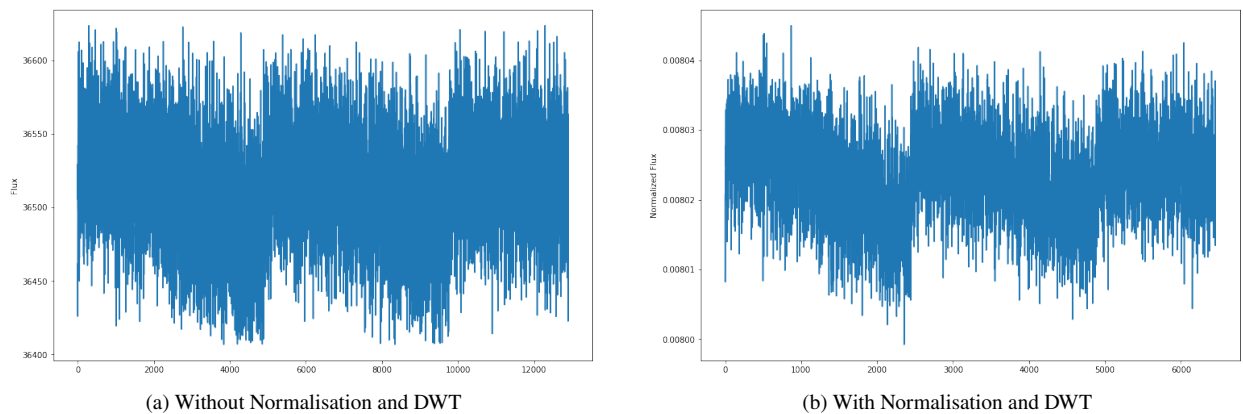

**Figure 3.** Sample lightcurve for class 0 (Eclipsing Binary)

classes, demonstrating the impact of the DWT on the light curves.

In addition to the light curves, we computed various statistical features from each individual curve. These features encompassed the mean, standard deviation, variance, kurtosis, and skewness, and were calculated from the original light curves prior to undergoing the DWT process. These statistical parameters were chosen as supplementary inputs to our model due to their discriminative properties, which became evident when examining the parameter space plots, as illustrated in Figure 2. The plots depict the relationship between the Mean and Standard Deviation, as well as the relationship between the Mean and Skewness. Notably, a visually discernible separating boundary, represented

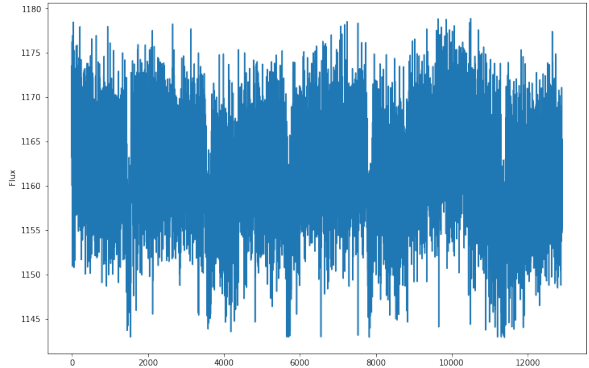 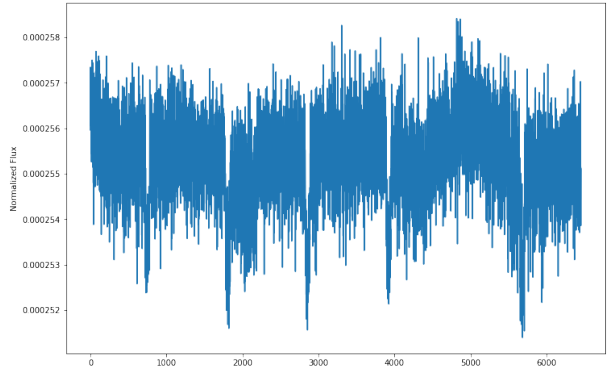

(a) Without Normalisation and DWT

(b) With Normalisation and DWT

**Figure 4.** Sample lightcurve for class 1 (Host Star)

by the black dotted line, emerged between the different classes separating a majority of points.

## 2.2 Model Architecture

The proposed model is designed to accept two different types of inputs in order to improve both accuracy and generalization. To accomplish this, we included two separate input layers in the model, each accepting a different type of input. These inputs are then combined at a concatenation layer, allowing the model to utilize the information from both inputs.

We name these input layers as "parameter layer" and "lightcurve layer" based on the type of input they take in. The parameter layer takes in the statistical features calculated earlier, while the lightcurve layer takes in the time series data of the light curves after applying DWT. **The complete model architecture is illustrated in Figure 5, providing an overview of each layer's precise dimensions and structure.**

### 2.2.1 Parameter Layer

**The statistical parameters calculated earlier serve as input to the parameter layer, which is connected to a fully connected neural network. This network comprises 5 input neurons that are connected to two fully connected layers, each containing 500 units. In a fully connected layer, every neuron in one layer is connected to every neuron in the subsequent layer. This means that each neuron in the preceding layer provides input to each neuron in the current layer, and each neuron in the current layer contributes to the output of every neuron in the succeeding layer. Within a fully connected layer, each neuron computes a weighted sum of the input values from the previous layer and applies an activation function, such as the rectified linear unit (ReLU), to generate an output value. This architectural design is widely used in neural networks and has been demonstrated to effectively capture non-linear relationships between the input and output vector spaces in various studies, including Khalifa et al. (2017); Nieto et al. (2019); Li et al. (2020).**
**To introduce non-linearity into the model, the ReLU activation function Agarap (2018); Ramachandran et al. (2017) defined as**

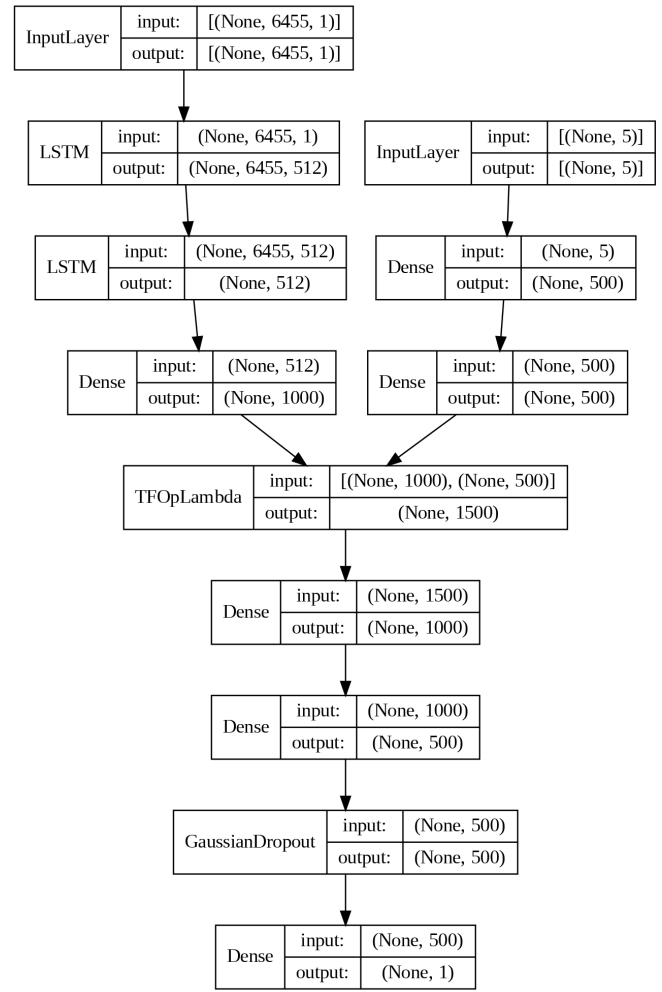

**Figure 5.** Model Architecture

$f(x) = max(0, x)$ **is employed in this layer. ReLU is a simple non-linear function that maps an input value to the maximum of 0 and the input value itself. In other words, it outputs 0 for negative input values and preserves the input value for non-negative or zero inputs. The effectiveness of ReLU in deep learning architectures has been demonstrated in studies such as Agarap (2018)** and Kulathunga et al. (2021). Its performance can be attributed to its non-saturating nature, which helps alleviate the vanishing gradient problem, and its ability to enable the network to learn sparse representations.

*2.2.2 LightCurve Layer*

The light curve layer is an independent input layer that is introduced to regularize the model and extract the inherent features of the light curves that the statistical parameters cannot capture. This layer is made up of LSTM units, which are known for their exceptional performance when it comes to sequential inputs Staudemeyer & Morris (2019a) such as time series data. Unlike their predecessors, RNNs, LSTMs have the property of long-term memory, which helps them to relate data points of a sequence over a longer duration, as discussed in studies such as Yu et al. (2019a) and Smagulova & James (2019). RNNs can only gaze back in time for a maximum of 10 timesteps Staudemeyer & Morris (2019b). This is because the signal being given back are either vanishing or exploding. Long-Short Term Memory Recurrent Neural Networks were developed to overcome this issue (LSTM-RNN) Mozer (1991); Hochreiter (1991).

In the proposed model, we use an LSTM layer with 512 units that take light curves as an input and generate a secondary sequence, which is then passed on to another set of 512 LSTM units. This is connected to a fully connected layer of 1000 units.

*2.2.3 Concatenation Layer*

The concatenation layer combines both the parameter layer and the lightcurve layer. It connects them to a fully connected layer of 1000 units, which is then connected to a layer with 500 units. This is followed by a Gaussian Dropout Srivastava et al. (2014) with a dropout rate of 0.2. The Gaussian dropout is a regularization technique that aims to prevent over-fitting and improves generalization by randomly dropping out network connections. This increases the training time but results in better generalization and prevents over-fitting Srivastava et al. (2014). The dropout layer then connects to the output layer, which has a single unit activated by a sigmoid Narayan (1997) function. This allows the output layer to produce a result between 0 and 1 which also represents the probability of the prediction.

**2.3 Training and Validation**

The proposed model is implemented using TensorFlow Abadi et al. (2015) and trained on Google Collaboratory, utilizing both an NVIDIA K80 GPU Carneiro et al. (2018) and an offline NVIDIA RTX 3060 GPU. To train the model, the dataset is divided into a $70:30$ **ratio, with** $70\%$ **of the data allocated for training and the remaining** $30\%$ **used for validation and testing. Including validation data is crucial for tuning the model's hyperparameters. The training process consists of** 1000 **epochs, and a stopping condition is incorporated to prevent overfitting** Caruana et al. (2001); Song et al. (2019); Ying (2019). **The training accuracy exhibits continuous improvement, while the validation accuracy reaches a saturation point and begins to oscillate, as illustrated in Figure** 6c. **A similar trend can be observed in the loss, refer to Figure** 7a7b **and** 7a **This inclusion of a stopping condition not only reduces computational requirements but also minimizes the risk of overfitting.**

**To optimize the model, Adam** Kingma & Ba (2014) **is chosen as the optimizer due to its robustness to noise and its adaptive learning rate properties. Additionally, Adam is a popular choice among practitioners in the field of deep learning because it incorporates momentum and is a variant of the AdaGrad optimizer, which facilitates quicker convergence.**

**Furthermore, since the model is designed for binary classification, the appropriate choice for the loss function is Binary Cross Entropy. Binary Cross Entropy is commonly used in binary classification tasks in machine learning. It quantifies the discrepancy between the predicted probability distribution of the binary output and the true probability distribution of the binary output. The Binary Cross Entropy loss function can be defined as:**

$$BCE(y, \hat{y}) = -\frac{1}{N} \sum_{i=1}^{N} [y_i \log(\hat{y}_i) + (1 - y_i) \log(1 - \hat{y}_i)]$$

**In this equation,** $y$ **represents a vector of true binary labels,** $\hat{y}$ **represents a vector of predicted probabilities for each binary label, and** $N$ **denotes the number of samples in the dataset. The BCE formula calculates the average loss across all the samples in the dataset. The first term in the square brackets corresponds to the loss when the true label is 1, while the second term represents the loss when the true label is 0. The negative sign is introduced to ensure the convexity of the loss function, which facilitates optimization using gradient descent methods.**

**Gradient descent is an optimization algorithm commonly employed to minimize the loss function of a model by iteratively adjusting the weights and biases, thereby improving the accuracy of predictions. The update rule for gradient descent can be expressed as:**

$$\theta_{t+1} = \theta_t - \eta \nabla L(\theta_t)$$

**Here,** $\theta_t$ **represents the vector of parameters at iteration** $t$, $\eta$ **denotes the learning rate that determines the step size at each iteration, and** $\nabla L(\theta_t)$ **represents the gradient of the loss function with respect to the parameters at iteration** $t$.

**Added Recently**

**In the case of Adam, similar principles apply. The parameters** $\theta_t$ **represent the vector of parameters at iteration** $t$. **The learning rate** $\eta$ **still determines the step size or the rate at which the parameters are updated. Additionally,** $\nabla L(\theta_t)$ **continues to represent the gradient of the loss function with respect to the parameters at iteration** $t$. **Therefore, the explanation can be adapted to Adam. The concepts of learning rate, parameter updates, and gradient calculations still hold true in the context of Adam optimization. The learning and decay rate of the optimizer has been fine-tuned to** $10^{-4}$ **through a process of experimentation involving various permutations and combinations within a similar range.**

**3 RESULTS AND DISCUSSION**

**In our study, we have developed a star classifier that incorporates wavelets and Deep Learning techniques to classify exoplanets and binary stars. Our proposed model demonstrates an accuracy of** $81.17\%$ **in correctly predicting the classes. The accuracy calculation is performed using the provided confusion matrix (see Table** 1)

**Another metric used to evaluate the performance of our model is the F1 score. The F1 score is a useful metric for evaluating the performance of our model. It takes into account both precision**

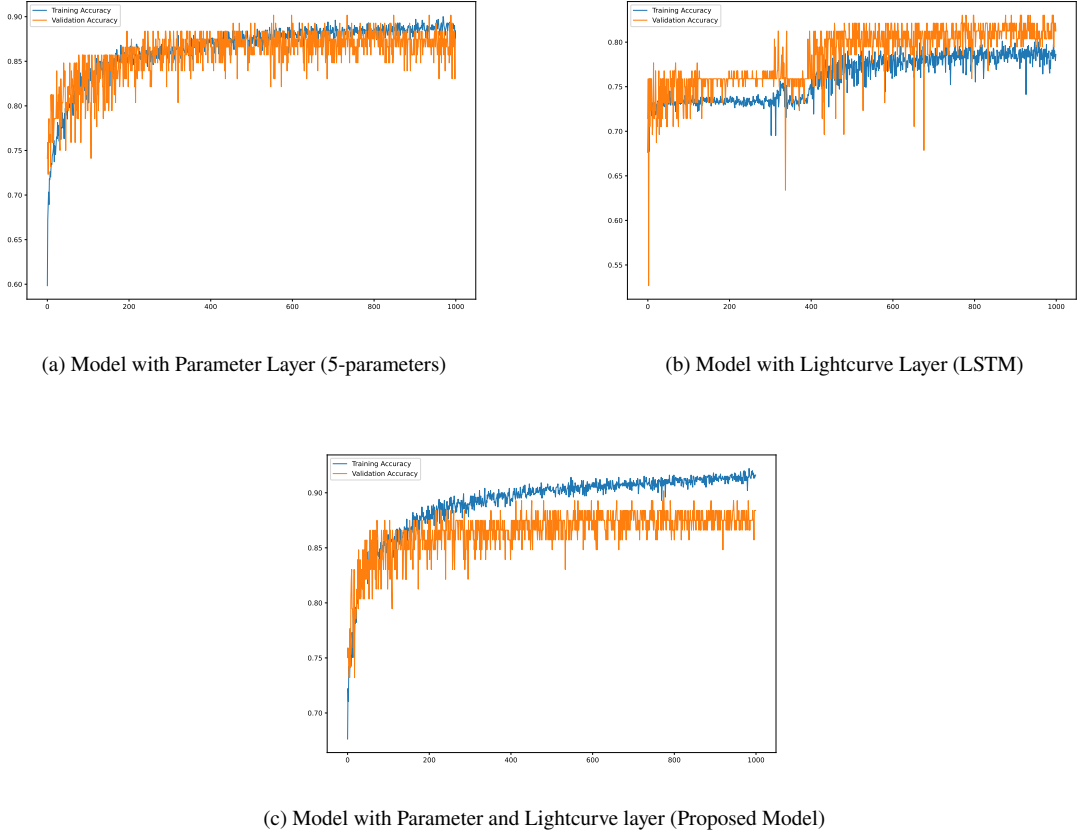

(a) Model with Parameter Layer (5-parameters)

(b) Model with Lightcurve Layer (LSTM)

(c) Model with Parameter and Lightcurve layer (Proposed Model)

**Figure 6.** Epochs vs Accuracy for different models

and recall, which are bounded between 0 and 1. Precision represents the false positive rate, indicating the rate of misclassified data for a particular class. Recall, on the other hand, represents the false negative rate. Analyzing Table 2, we observe that the recall for class 0 is relatively lower compared to that of class 1. This suggests that more exoplanets are being predicted as eclipsing binaries, while fewer binaries are being predicted as exoplanets for the proposed model. The F1 score offers several advantages over accuracy. Firstly, it is more sensitive to imbalanced datasets, where accuracy can be misleading. For instance, a model that consistently predicts the majority class may achieve high accuracy, but it may fail to correctly identify the minority class. In such cases, the F1 score will provide a lower value, indicating the model's performance accurately. Additionally, the F1 score strikes a balance between precision and recall, which is crucial in different applications. Some scenarios prioritize high recall (minimizing false negatives), while others prioritize high precision (minimizing false positives). The F1 score considers both factors, making it a more comprehensive measure of model performance.

Additionally, in the context of binary classification, the ROC curve serves as a visual representation of the algorithm's performance. It depicts the true positive rate (sensitivity) against the false positive rate (1 - specificity) at various classification thresholds. The ROC curve aids in assessing the trade-off between correctly identified positive instances and incorrectly identified negative instances. The area under the ROC curve (AUC) is a valuable metric that summarizes the overall performance of the classifier. It quantifies the classifier's ability to distinguish between positive and negative instances across all possible thresholds. The AUC provides a single-number summary, allowing for easy comparison between different classifiers. The ROC curve is essential in classification tasks as it enables us to visualize the relationship between the true positive rate and the false positive rate, aiding in the selection of an appropriate classification threshold. Furthermore, the AUC serves as a comprehensive measure of classifier performance, capturing the classifier's discriminatory power across all possible thresholds.

According to Figure 8c, the proposed model achieves an AUC value of $0.81$. In an ideal scenario, the AUC value would be 1, indicating perfect classification. One of the advantageous properties of ROC curves is their insensitivity to changes in class distribution. This means that even if the proportion of positive to negative instances varies in a test set, the ROC curve remains unaffected, unlike accuracy, which can be influenced by such changes.

To further evaluate the importance of our input data, we compared our proposed model against two other models. The first model solely takes the processed lightcurve as input, while the second model only considers the 5 parameters as input. Both models were trained using a similar approach for 1000 epochs on the same training data and subsequently tested on the same test data as our proposed model. This was done to ensure an unbiased analysis of the results.

Upon comparing our proposed model with the two alternative models, it is evident that they exhibit lower validation accuracy

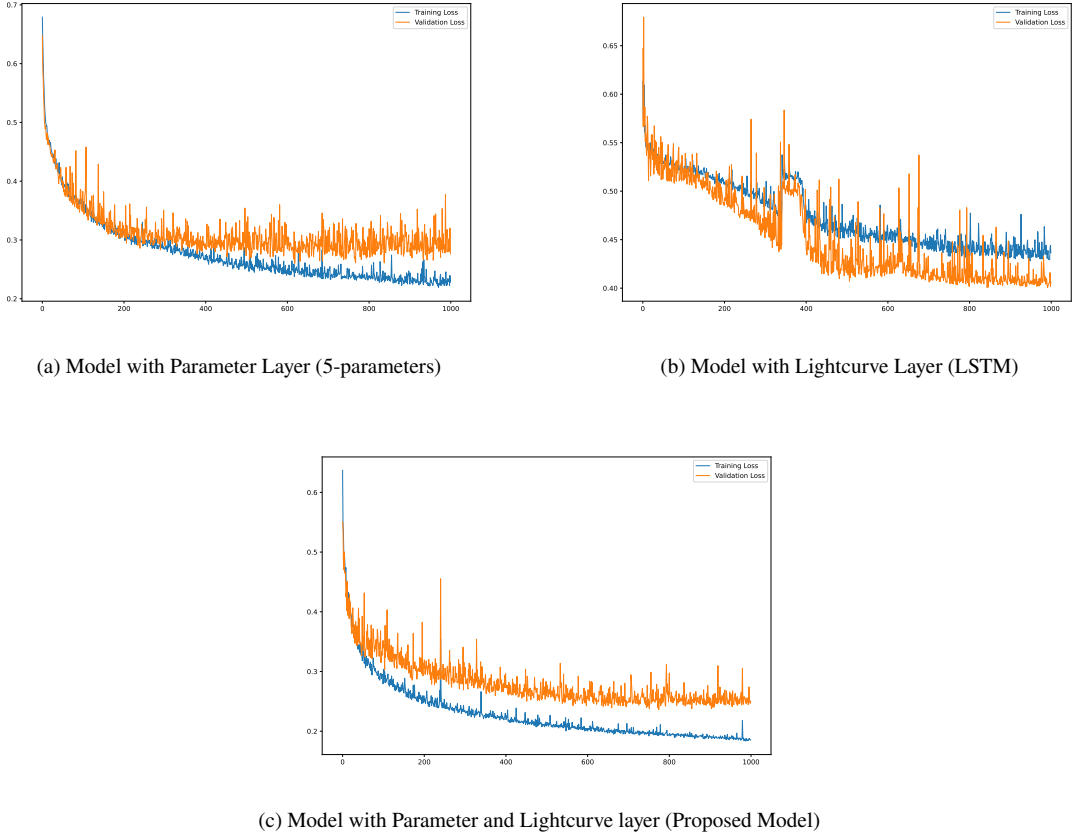

(a) Model with Parameter Layer (5-parameters)

(b) Model with Lightcurve Layer (LSTM)

(c) Model with Parameter and Lightcurve layer (Proposed Model)

**Figure 7.** Epochs vs Loss for different models

**Table 1.** Confusion Matrix for every model

|  | Class 0 | | | Class 1 | | |
|---|---|---|---|---|---|---|
|  | 5-parameters | LSTM | Proposed Model | 5-parameters | LSTM | Proposed Model |
| Class 0 | 171 | 203 | 185 | 42 | 74 | 60 |
| Class 1 | 20 | 55 | 30 | 213 | 178 | 203 |

**Table 2.** Precision, Recall and F1 score for every model

|  | Precision | | | Recall | | | F1 Score | | |
|---|---|---|---|---|---|---|---|---|---|
|  | 5-parameters | LSTM | Proposed Model | 5-parameters | LSTM | Proposed Model | 5-parameters | LSTM | Proposed Model |
| Class 0 | 0.90 | 0.79 | 0.88 | 0.70 | 0.83 | 0.72 | 0.78 | 0.81 | 0.79 |
| Class 1 | 0.74 | 0.81 | 0.75 | 0.91 | 0.76 | 0.90 | 0.82 | 0.79 | 0.82 |
| Average | 0.82 | 0.80 | 0.81 | 0.81 | 0.80 | 0.81 | 0.80 | 0.80 | 0.80 |

scores, as depicted in Figure 6a and Figure 6b. However, when evaluating their accuracy on the same test data, Table 1 indicates an accuracy of 80.33% for the 5-parameter-only model, while Table 1 reveals an accuracy of 79.70% for the LSTM-only model. These results clearly demonstrate that our proposed model surpasses the traditional model trained solely on the 5 statistical parameters extracted from the lightcurves, as well as the LSTM-based model that directly takes the lightcurves as input.

This superiority is further supported by the ROC curves of the two models, presented in Figure 8a and Figure 8b. The AUC value for the 5-parameter model is 0.806, while the AUC for the lightcurve-only model is 0.796. Both of these values are lower than the AUC of our proposed model, affirming the superior perfor-

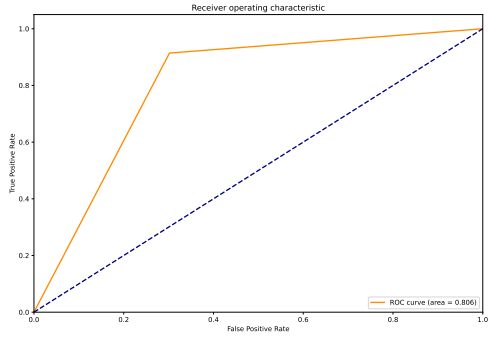

(a) Model with Parameter Layer (5-parameters)

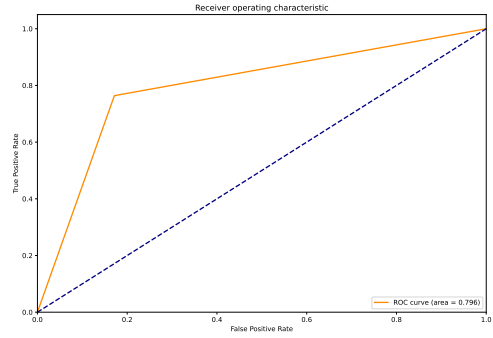

(b) Model with Lightcurve Layer (LSTM)

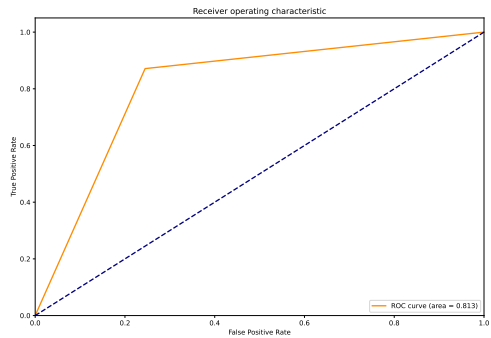

(c) Model with Parameter and Lightcurve layer (Proposed Model)

**Figure 8.** ROC curve for different models

mance and generalizability of our approach in the classification task.

## 4 CONCLUSION

The current study introduces a novel approach that combines a wavelet-enabled Long Short-Term Memory (LSTM) model with key statistical features to accurately classify binary eclipses and exoplanet stars. A notable advantage of our algorithm is its ability to make predictions based solely on light curve data. In many cases, distinguishing between similar transits poses a significant challenge. However, our approach effectively tackles this challenge by carefully selecting and utilizing the appropriate feature space. Our algorithm achieves an impressive accuracy of 81.17%, showcasing substantial progress in the classification process. This achievement is particularly noteworthy given the inherent difficulties arising from the high similarity in light curve data, which poses challenges in identifying distinctive features. Moreover, our model achieves an average F1 score of 0.80, indicating a well-balanced performance in terms of precision and recall. Overall, our study presents a promising advancement in the field of classification, addressing the challenges associated with similar transit patterns and demonstrating the efficacy of our approach in accurately categorizing binary eclipses and exoplanet stars based on light curve data.

## ACKNOWLEDGEMENTS

This research uses Lightkurve, a Python package for analyzing Kepler and TESS data Collaboration et al. (2018). The TESS mission is funded by the Science Mission directorate of NASA. This study utilized ASTROPY, a community-developed Python core package for astronomy Price-Whelan et al. (2018). This study also employed astroquery, a Python package for astronomical web-querying Ginsburg et al. (2019).

## DATA AVAILABILITY

The paper contains TESS mission data that are publically accessible from the Mikulski Archive for Space Telescopes (MAST). The data and code used can be found here https://github.com/amanasci/hs-eb-classifiction

This paper has been typeset from a T<sub>E</sub>X/L<sup>A</sup>T<sub>E</sub>X file prepared by the author.